# Learning the nature of viscoelasticity in geologic materials with MCMC

Ron Maor[a], Lars Hansen[b], Douglas Jerolmack[a], and David Goldsby[a]

**Rock and ice are ubiquitous geologic materials. While apparently solid, they also exhibit fluid behavior under stress – a property termed viscoelasticity. Viscoelastic convection of Earth's mantle drives tectonic plate motion with consequences for earthquakes and sea-level rise, while viscoelastic deformation of ice controls glacier flow and the flexure of icy moons. For crystalline materials, "flow laws" describing bulk rheology can be derived from understanding microstructural dynamics such as crystal-defect migration. Common geologic materials like ice and olivine have grain sizes and crystal orientations that evolve with strain; this complexity precludes a first principles approach. Here we use a Bayesian inference method to learn the connection between microstructure and flow in ice and olivine, from fits to experimental data of these materials undergoing steady-state deformation and forced oscillations. We demonstrate that this method can constrain a nonlinear viscoelastic model for each material, that is capable of capturing both steady and transient dynamics and can also predict dynamics for data it was not trained on. Our results may improve geodynamic models that rely on parameterized constitutive equations, while our approach will be useful for experimental design and hypothesis testing.**

Nonlinear Viscoelasticity | MCMC | Ice | Olivine | Attenuation

**V**iscoelastic materials show markedly different behaviours when subjected to external forces on varying timescales. On short time scales, viscoelastic materials respond with instantaneous deformation, known as elasticity. On long timescales, however, viscoelastic materials may show viscous flow behavior, continuing to deform under the applied stress. Rocks and ice demonstrate viscoelastic properties. In thermal convection, lithospheric rocks show more elastic behavior compared to those in the sublithospheric mantle, which behave more like a viscous fluid (1). The dynamics of ice sheets are influenced by gravitational forces, tides, and wave action, each operating on different timescales and contributing to both short-term elastic effects and long-term non-Newtonian viscous flow (2). Viscoelastic models are composed of constitutive relations between stress and strain, combining both elastic and viscous responses. When the strain of the materials is directly proportional to the applied stress, the constitutive relations are linear, and the material is considered linear viscoelastic. When the relationship between stress and strain is nonlinear, the material is considered nonlinear viscoelastic. Nonlinear viscoelasticity is associated with a wide variety of large-scale phenomena (Figure 1). For example, relative sea-level rise controlled by the response of the crust and mantle to melting ice sheets, known as Glacial Isostatic Adjustment (GIA), is better fit by models that use nonlinear viscoelasticity than by models that use linear viscoelasticity (3). In addition, the dynamics of fault slips during seismic cycles are influenced by nonlinear viscoelastic properties (4). Other examples include modeling of mantle convection in the Earth and icy satellites using non-Newtonian viscosity (5, 6). The nonlinear behavior inferred for all of these large-scale systems is motivated by and consistent with small-scale laboratory experiments, that routinely observe power-law relations between stress and strain (e.g. 7–9).

The nature of viscoelasticity lies in the molecular properties and internal structure of the material. Viscoelasticity is attributed to factors such as grain boundary interactions, temperature, pressure, and defects in the crystalline structure. Defects can accommodate strain through different deformation mechanisms, such as diffusion creep, grain boundary sliding, and dislocation creep (10–12). Laboratory experiments are able to simulate pressure and temperature conditions found in natural settings, and deform material samples under controlled strains/stresses to activate different deformation mechanisms. Moreover, tools like Electron Backscatter Diffraction (EBSD) enable the observation of microstructure changes in the sample before and after deformation (13). Studies suggest that deformation

## Significance Statement

Rocks in Earth's interior and ice behave elastically on short timescales, but flow on geologic timescales. This 'viscoelastic' behavior underlies the mantle convection that drives Earth's plate tectonics, the flow of glaciers under a changing climate, and the squeezing and stretching of moons under tidal forcing. We know that the evolving shape and structure of tiny grains modulates the bulk response of rocks and ice to stress, but it is not always clear how for such complicated materials. Here we use machine learning to approximate the viscoelastic behavior of ice and the mantle mineral olivine from laboratory experiments, and deduce how their deformation depends on the grain-scale processes. These flow laws are physically sensible and have predictive power, which may help to improve models for planetary dynamics and ice flow.

Author affiliations: [a]University of Pennsylvania; [b]University of Minnesota

[2]To whom correspondence should be addressed. E-mail: ronmaor@sas.upenn.edu

by dislocations is prevalent at temperatures and pressures typically experienced by geological materials, and therefore is expected to play a key role in numerous large-scale phenomena. For example, it is believed that a significant portion of the mantle deforms by dislocation creep (e.g. 14–16), and that dislocation creep plays a crucial role in the flow of ice masses and glaciers (17, 18). Two microstructural parameters that are often associated with the mechanics of dislocation creep in polycrystals are the average grain size and Crystallographic Preferred Orientation (CPO), or texture. Dynamic recrystallization is a process by which new grains form, with finer grain size and different orientations, under strain. Shear strain during deformation can also cause grains to rotate and achieve more uniform orientations. Dynamic recrystallization has been observed in both ice and olivine, and shown to significantly affect to the mechanical properties of these materials (19–21). At the planetary scale, the effects of grain size and texture have been demonstrated in seismic wave propagation, tectonics, and mantle convection (22–24).

Deformation experiments quantify the connection between microstructural properties and macroscopic deformation by formulating 'flow laws': mathematical relations between stress, strain, and strain rate. To model the full spectrum of viscoelastic behavior, three components need to be incorporated: instantaneous elastic deformation, a transient component, and steady-state viscous flow (the 'flow law'). The simplest type of viscoelastic model that contains these three types of deformation and is popular in geodynamics is the Burgers model. It is a mechanistic rheological model that can be visualized as a combination of two elastic springs and two viscous dashpots. The Burgers model has been used to explain multiple phenomena, such as the tidal bending of glaciers (25), tidal heating in Saturn's icy moon Enceladus (26), GIA signals (27, 28), and the brittle-ductile transition in the Earth's crust (29). The parameters of the model are the moduli of the springs and the viscosities of the dashpots, which can be inferred in a Bayesian framework by running Markov Chain Monte Carlo (MCMC) simulations with laboratory data (e.g. 30–32). Deformation by dislocations has been extensively studied in laboratory settings on ice and olivine (10, 33–35), and has been characterized using a nonlinear 'flow law', relating the stress and strain rate through a power law. To combine the understanding of dislocation creep mechanisms from lab experiments with the geodynamics framework, the viscoelastic Burgers model can be adjusted to include the nonlinear flow law. Then, deformation data can be used with MCMC methods to assess the parameters of the improved model and test its ability to fit the data. Such an approach was previously used with olivine (36, 37).

Steady-state flow laws of geological materials are much more established and better constrained than the time-dependent transient ones. The experimental conditions needed for measuring the transient phase are more challenging to achieve, and the small strains involved make it difficult to conduct precise and reproducible measurements. During the transient phase, the microstructure undergoes considerable changes, which increases the uncertainty of the deformation dynamics. Complicating matters further, the time-dependent deformation during the transient phase is often composed of two different types of strains occurring simultaneously: a non-recoverable strain (transient creep) and a recoverable strain with energy dissipation (anelasticity) (38). Anelasticity can be measured in forced oscillation experiments (39, 40), and transient creep is measured in stress relaxation or stress drop experiments (41, 42). Mechanical models such as the Burgers model do not incorporate microstructural dynamics, and they are usually tested against one type of experimental data (creep experiments or forced oscillations). Moreover, the nonlinear viscoelasticity has never been incorporated into the modeling of forced oscillations, which contributes to the lack of understanding of its influence on the attenuation spectrum at low frequencies relevant to tidal forcings. There is a need for a comprehensive nonlinear viscoelastic model that can be used for a variety of datasets, that would bridge the gap between observations in the lab and geodynamic modeling.

Here we incorporate microstructural dynamics into the Burgers model framework and use the same model archetype to fit data from dislocation creep experiments on ice and olivine. We propose a model that captures the entire viscoelastic spectrum and fits data from recent experiments on both materials. To infer the parameters with efficient convergence and minimal tuning, we use a type of MCMC method that is a recent extension of the Hamiltonian Monte Carlo (HMC) algorithm. We employ data from constant strain rate and dynamic oscillation experiments, testing the versatility of the model. Additionally, we extrapolate the optimized model to fit experiments that were not part of the MCMC simulations and make predictions for nonlinear viscoelastic effects that can be measured in future experiments.

## Theoretical Background

The complex nature of deformation mechanisms in olivine was previously studied using MCMC methods. Jain et al. (51) examined a wide array of published experimental data on olivine and used MCMC techniques to better constrain the parameters of composite flow laws, focusing on various steady-state creep mechanisms without transient components. Masuti et al. (36) performed MCMC analysis on experiments involving constant strain rates and strain rate steps conducted by Chopra et al. (52, 53) on natural dunites. They employed a nonlinear Burgers-type model for the analysis, focusing on capturing both the transient and steady-state creep behavior of the deformation. Masuti et al. (36) introduced nonlinearity to both the steady-state and transient creep, but the dynamical evolution of the microstructure was not explicitly included. In this study, the model formulation uses a nonlinear Burgers-type framework, where the effects of CPO and evolving grain size are incorporated. These two microstructure parameters are routinely measured in the lab and provide observable insight into the dynamics of the deformation.

The mechanical circuit of a Burgers model is composed of a Maxwell model and a Kelvin-Voigt model connected in series (Figure 2). The Maxwell model has a spring and a dashpot connected in series, representing the instantaneous elastic deformation and steady state viscous flow respectively. The Kelvin-Voigt model has a spring and a dashpot connected in parallel, capturing the anelastic transient component of the deformation. When elements are connected in series, the applied stress ($\sigma$) is the same on each element (similar to current in an electrical circuit), and the total strain ($\varepsilon$) is distributed between the elements (similar to voltage in an electrical circuit). In a parallel configuration, the total strain is the same in each element, while the stress is distributed. Consequently, the macroscopic strain during the deformation of a Burgers model is the sum of the strains in the Maxwell and Kelvin-Voigt circuits. To align the Burgers model rheology with experimental datasets from the deformation of ice and olivine in the dislocation creep regime, we introduce nonlinearities and account for additional microstructural dynamics. In the following, we go over the main considerations of our model; for the full set of the model equations, refer to *SI Appendix*, Model equations.

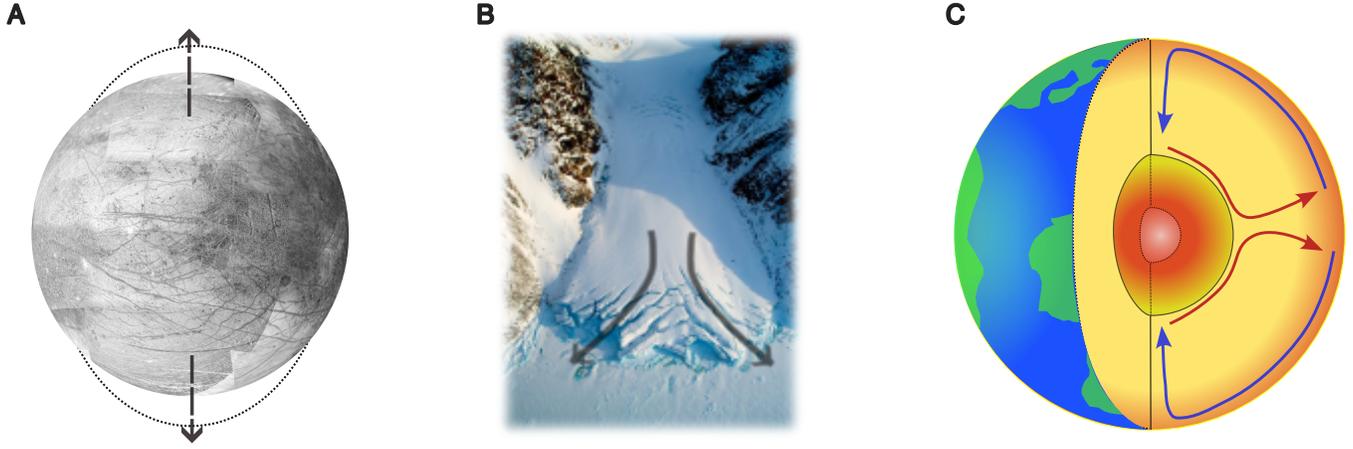

**Fig. 1.** The role of viscoelasticity in geodynamic phenomena. (A) The tidal response of the icy shell of Jupiter's moon Europa is modeled using viscoelastic models (e.g. 43, 44). Picture of Europa from NASA Image and Video Library (45) (B) Viscoelastic properties are incorporated in studies of outlet glacier flows and ice shelf flexure (e.g. 46, 47). Picture of a small valley glacier exiting the Devon Island Ice Cap, from NASA Image and Video Library (48). (C) Models of mantle flow and thermal evolution use viscoelasticity to represent the time-dependent deformation (e.g. 49, 50).

**Steady State Viscous Flow.** The steady state dislocation creep (and other mechanisms) in ice and olivine is represented by an Arrhenius-type 'flow law' (e.g 33, 34, 54). Here we use a similar form to represent the strain of the Maxwell dashpot:

$$\dot{\varepsilon}_{M_d} = \left(\frac{\sigma}{F}\right)^n d^{-p} e^{Q_M\left(\beta - \frac{1}{RT}\right)} \quad [1]$$

where $\sigma$ is the principal axial stress and $\dot{\varepsilon}_{Md}$ is the strain rate of the dashpot, $R$ is the universal gas constant, $T$ is the material temperature, $Q_M$ is the activation enthalpy, $d$ is the average grain size, $n$ is the stress exponent, $p$ is the grain size exponent, and $\beta = \ln A/Q_M$ (where $A$ is the material constant or pre-exponential factor in relevant literature). It is important to note that $\beta$, $n$, $p$ and $Q_M$ depend the material and the particular deformation mechanism. The parameter $F$ is a measure of the geometric softening; i.e., how the mechanical strength decreases during plastic deformation as CPO evolves. When $F = 1$ it indicates that there is no CPO and the polycrystalline material is isotropic. The value can decrease towards the lower limit of 0 based on the strength of the dominant slip system.

**Transient Component.** Transient creep refers to the initial, time-dependent deformation that occurs before steady-state flow is established. It becomes particularly significant in situations where there is a notable contrast in rheological properties between low-strain transient deformation and high-strain steady-state deformation. Despite its importance, substantial uncertainty remains surrounding the micromechanical processes associated with transient creep, mainly due to experimental constraints. The limited duration of the transient phase, the constantly changing microstructure, and the rapid changes in stress or strain rates make transient creep experiments more challenging.

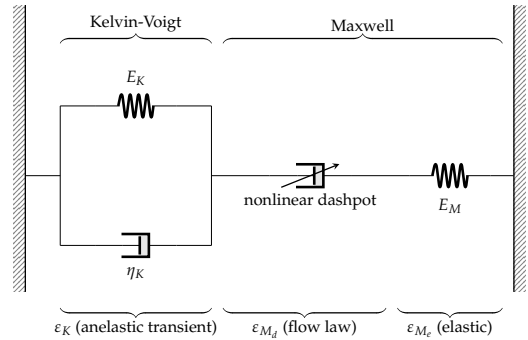

**Fig. 2.** The nonlinear Burgers model of this study. It has the same circuit as the linear Burgers model, except that the Maxwell dashpot represents a nonlinear flow law with evolving microstructure.

Here we rely on recent experiments and theory to evaluate an appropriate formulation for the transient part of our model, focusing on dislocation dynamics. A single dislocation in the crystal lattice produces a stress field. When dislocations move, they interact with each other elastically and alter the overall stress distribution. As a result, plastic deformation is not solely determined by the applied stress but is also influenced by the backstress $\sigma_b$ (or internal stress) resulting from these interactions. The creep rate in this scenario is assumed to result from the effective stress $\sigma_e$, which is the difference between the externally applied stress and the backstress $\sigma_e = \sigma - \sigma_b$ (41, 55).

To measure the backstress and its associated microstructure, (41) performed a set of stress reduction experiments on olivine. In their experiments, the sample was kept under constant load process until a steady-state strain rate was attained. Following this, the load was promptly reduced and held at a lower value while monitoring the strain. Assuming a constant backstress during the reduction (i.e., rapid reduction), the sign of the measured plastic strain rate after the load reduction can provide information about the backstress. The magnitude of the stress reduction that results in zero strain rate is the backstress. They observed strain hardening and concluded that the transient deformation is being controlled by dislocation glide.

(56) suggested a theoretical model of dislocation dynamics to explain steady state and transient creep. The model introduces the following flow law:

$$\dot{\varepsilon} = A(T)\rho \sinh\left(\frac{\sigma_e}{\sigma_{\text{ref}}(T)}\right), \qquad [2]$$

where $A(T)$ is a pre-factor that depends on the temperature, $\sigma_{\text{ref}}(T)$ is a reference stress that depends on the temperature, and $\rho$ is the dislocation density. The model showed a good fit for the transient creep experiments performed by (41). At low applied stress, the backstress is able to balance it and the resulting effective stress is low. In that case, the sinh function in Eq. (2) can be approximated as a linear function, and the strain rate is linearly dependent on the stress. The linearity is also observed experimentally.

In light of these studies, we propose that the transient component of our model can be effectively modeled by a singular linear Kelvin-Voigt element, similar to the one already integrated into the Burgers model. Conceptually, the applied stress is distributed between the dashpot and the spring in the Kelvin-Voigt circuit. This distribution results in a low effective stress on the dashpot, justifying the linear approximation for Eq. (2).

**Microstructure Evolution.** To incorporate evolving grain-size and CPO, we follow the parametrizations outlined in (19). We evolve the average grain size $d$ and CPO factor $F$ using phenomenological relaxation equations:

$$\frac{d}{dt}d = \dot{d} = \frac{\dot{\varepsilon}(d_{ss} - d)}{\varepsilon_{c1}} \qquad [3]$$

$$\frac{d}{dt}F = \dot{F} = \frac{\dot{\varepsilon}(F_{ss} - F)}{\varepsilon_{c2}} \qquad [4]$$

where $\dot{\varepsilon}$ is the *total* macroscopic strain rate, $d_{ss}$ is the average grain size at steady-state, $F_{ss}$ is the CPO factor at steady-state, $\varepsilon_{c1}$ is a parameter controlling the relaxation time represented by a characteristic strain required to approach the recrystallized grain size (57), and $\varepsilon_{c2}$ is the critical strain for CPO development.

## Results

**MCMC with ice Data.** The ice data in the MCMC runs are from the experiments done by (58). Polycrystalline ice samples were deformed at 263 K using a high-pressure gas-medium apparatus, which allowed dislocation creep at relatively high stresses without fracturing. The samples were subjected to uniaxial compression at a nominally constant strain rate until a steady-state flow stress was achieved. Dynamic recrystallization with grain size reduction and CPO development were observed in the samples. Figure 3 shows the experimental setup and the stress-strain curves after the MCMC optimization. The nonlinear model solution using the mean parameter values from the MCMC simulations shows good agreement with the experimentally measured data. Retrodiction curves, representing random samples from the posterior chains, are also presented in grey within the figure to visualize the associated errors.

The similar shape of all the curves indicates comparable dynamics. A peak stress occurs at around 3% strain, followed by a drop towards a steady-state value. This behavior is explained by the evolving microstructure, which works to soften the material after reaching a certain strain (58). In addition to the fact that our model successfully captures the dynamics of the experiments, we can use other metrics to infer its validity. In our Bayesian framework, the uncertainty of each parameter is quantified using a probability distribution (the prior). If the resulting posterior distribution of a parameter after the run converges to a narrow range that aligns with our expectations from other experiments or first principles, it increases the confidence in the model. Figure 4 shows the prior and posterior distributions for the experiment in Figure 3B. Two parameters that showed good agreement with expected values are the stress exponent $n$ and the critical strain for grain size evolution $\varepsilon_{c1}$. Notice that both started with an imposed wide and uniform distribution, and each converged to a narrow distribution with $n \approx 4$ and $\varepsilon_{c1} \approx 0.03$. The two parameters with the highest uncertainties belong to the Kelvin-Voigt elements, as can be seen in Figure 4(C-D). These two parameters showed convergence. This result is satisfying since it demonstrates the ability of the model to capture the dynamics of the experiment uniquely with this mechanistic formulation. The other posterior distributions and their statistics are included in *SI Appendix*, Ice (Fig. S1-S2, Table S2).

Importantly, the model is able to effectively capture the dynamics of deformation from the initial stages through the peak stress. This is intriguing because the peak stress of these experiments occurs relatively quickly (within a few minutes), meaning there are not many data points during this transient phase. Nonetheless, the model shows a good fit to the data from the initial loading and throughout each experiment. Indeed, the faster experiment shows larger errors around the peak stress (Figure 3B).

**MCMC with Olivine Data.** The experimental data of olivine in this study are from a forced oscillation experiment, wherein a sinusoidally-varying axial load was applied to the sample. For viscoelastic materials, the phase of the resulting strain response of the sample will lag behind the phase of the applied stress. Attenuation (also called the internal friction) in the sample can be calculated as the tangent of the angle of this phase lag ($tan\delta$). In the experiment referenced in this study, polycrystalline olivine samples were deformed under high-pressure, high-temperature conditions in the Deformation-DIA apparatus (59, 60). The samples underwent forced oscillations at a period of 300 seconds, at a temperature of 1423 ±50 K. Microstructural analysis of the samples before and after forced oscillations yielded no noticeable change in CPO strength. Therefore, the effect of CPO was not incorporated in the model. Grain size evolution was incorporated.

Figure 5 presents the results of the MCMC runs for the stress parameter. The mean values of the posterior distribu-tions, depicted by the dashed blue curve, effectively capture both the amplitude and phase of the signal. Retrodiction curves (grey lines) suggest that the predominant source of variance in the posteriors is associated with fitting the amplitude (*SI Appendix*, Olivine shows the full statistics of the parameters after the runs; see Table S3). The mean value of the stress exponent is 3, suggesting dislocation-based deformation mechanism (Fig. S5A). Interestingly, the value of the Kelvin-Voigt spring constant $E_K$ shifted from a uniform distribution prior to a narrow normal distribution posterior (Fig. S5C). This suggests the ability of the model to constrain the parameters of transient creep using attenuation experiments.

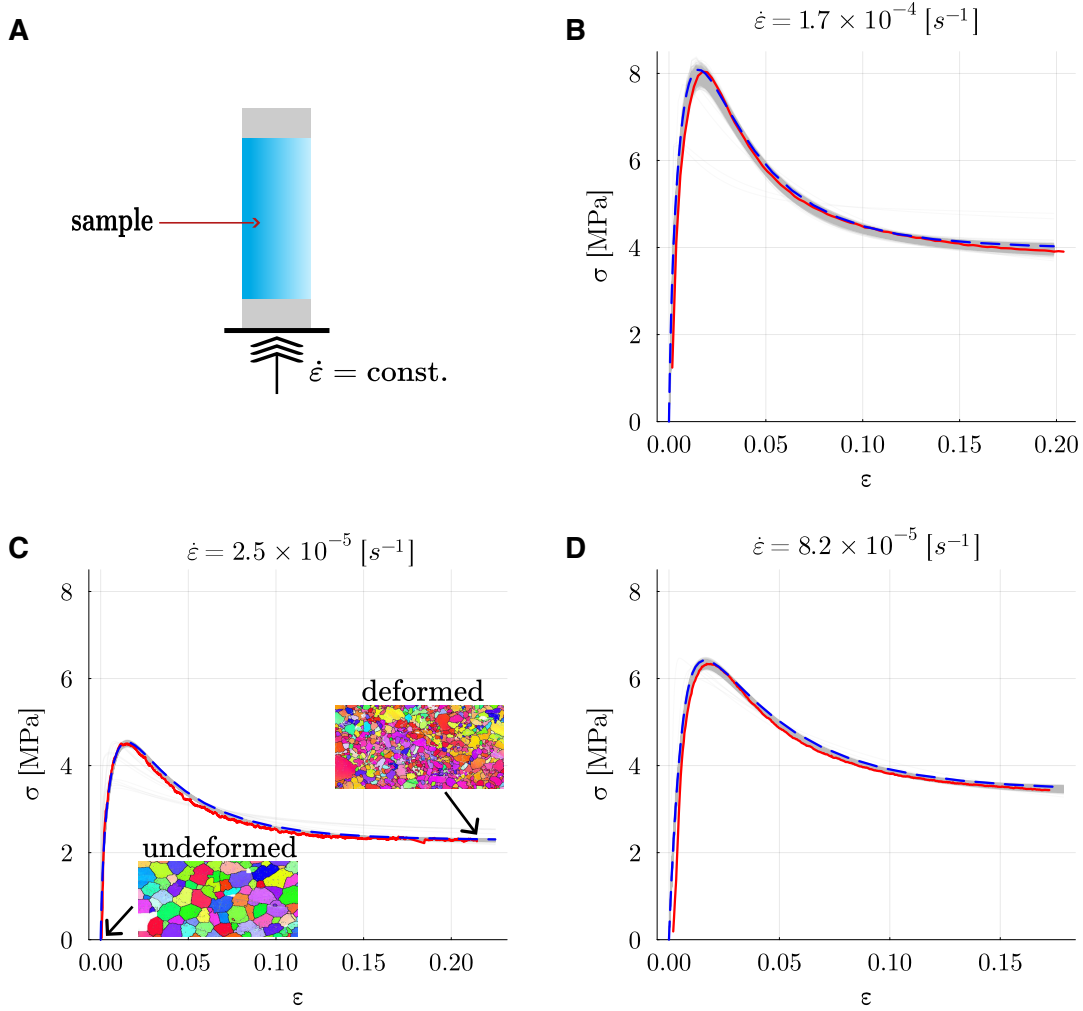

**Fig. 3.** (A) The constant strain rate ice experiment: the sample is deformed using a stepper motor that pushes the sample from the bottom at a constant rate. (B-D) Stress-strain curves from MCMC runs, along with data from (58). The experiments are referred to as PIL40, PIL72, and PIL105 in the original text. The red curve is the original experimental data set. The blue dashed curve is the nonlinear Burgers model solution with the mean value of the parameters from the posterior. The grey curves are retrodiction plots of 1000 randomly sampled parameter sets from the posterior. The temperature in the numerical solutions was set to 260 K, and the strain rates given by the experimental data are shown at the top of each curve. Panel (C) includes an EBSD analysis of one sample from (58), showing dynamic recrystallization. The colors represent the orientation of each grain. Notice that the deformed sample has more (and smaller) grains, and the orientations are more uniform.

**Predictions.** The demonstrated capability of the nonlinear model to achieve a good fit for both constant strain rate and forced oscillation experiments prompts the question of whether it can extrapolate results from one experimental type to another. To investigate this question, we use the optimized values from the MCMC runs on ice at a constant strain rate to reproduce the attenuation spectrum measured in a different experiment by (61). Their experiments were performed using a commercial servomechanical actuator operating under ambient pressure to apply a low-frequency cyclic stress on top of a median stress. Uniaxial loading was employed with a median stress ($\sigma_m$) of 1 MPa and a cyclic amplitude ($\sigma_0$) ranging from 0.05 to 0.28 MPa. Each sample was loaded under median stress for 1-2 hours until reaching steady-state creep, after which the periodic loading was initiated. Their findings suggested that the attenuation spectrum is insensitive to grain size and exhibits a slight nonlinearity, indicating amplitude dependence. The nonlinearity observed in the study provides a valuable benchmark for comparing with outcomes derived from our nonlinear model. Using the model, we numerically replicate the experiments and generate attenuation spectra for various stress amplitudes. We also replicate the nonlinearity tests (Fig. S3). This numerical approach offers advantages, as it is not bound by the physical constraints inherent in the laboratory setting and limitations of the apparatus. This enables exploration of additional stress amplitudes beyond those employed in the original study, or any other attenuation study on ice to date. The steps taken to use the optimized model as a numerical material analogue and produce the attenuation spectrum are described in *SI Appendix*, Ice Attenuation Predictions.

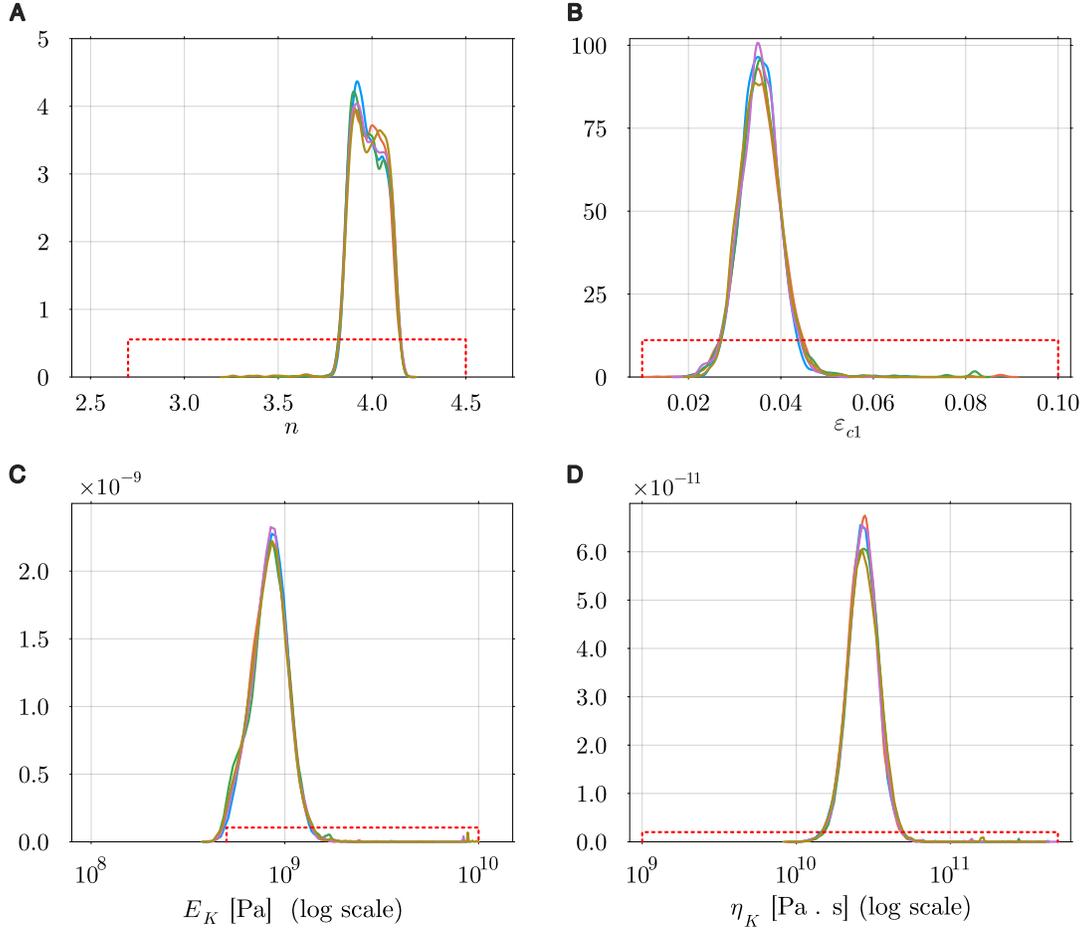

**Fig. 4.** Priors and posterior distributions of the MCMC runs of the PIL40 ice experiment of (58) (Figure 3B). Prior distributions are represented by red lines, according to Table 1, and the posterior distributions of the seven chains are the narrower distributions. Notice that the y-axis values are normalized so that the area under each curve sums to 1. (A) Stress exponent $n$. (B) Critical strain for grain size evolution $\varepsilon_{c1}$. (C) Spring modulus of the Kelvin element $E_K$. (D) Viscosity of the Kelvin dashpot $\eta_K$.

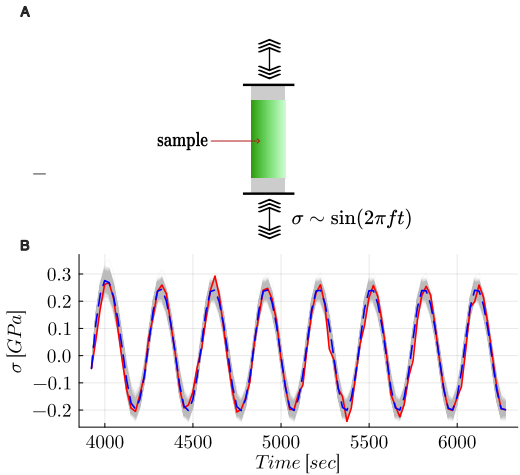

**Fig. 5.** Experimental setup and MCMC results for the olivine data. (A) The olivine sample was deformed using axial forced oscillations in the Deformation-DIA apparatus. (B) Stress data accompanied by the corresponding numerical results. The red line denotes the experimental data, while the dashed blue line represents the numerical fit with the mean values of the posterior distributions. The grey curves are retrodiction plots of 1000 randomly sampled parameter sets from the posterior.

Figure 6 shows the attenuation spectrum computed for four distinct steady-state creep regimes. The experiments conducted by (61) are numerically replicated by the blue curve. Notably, there is a scarcity of relevant data points in the experimental data, hindering a comprehensive comparison of the attenuation pattern. Nevertheless, the figure suggests a potential similarity in the overall attenuation pattern, characterized by a Debye peak and a power-law background signal at lower frequencies (sometimes reffers to as high-temperature background). Leveraging the freedom to examine many conditions with the model, we extended the attenuation calculations to elevated median stress levels of 1.5, 2, and 4 MPa to explore the nonlinear effects. Figure 6 illustrates that the attenuation curves for elevated stress levels exhibit an increase in attenuation and the presence of nonlinear characteristics. Notably, the effect of power law in the low frequencies intensifies at elevated stresses, a trend most noticeable at 4 MPa, where the slope of the curve

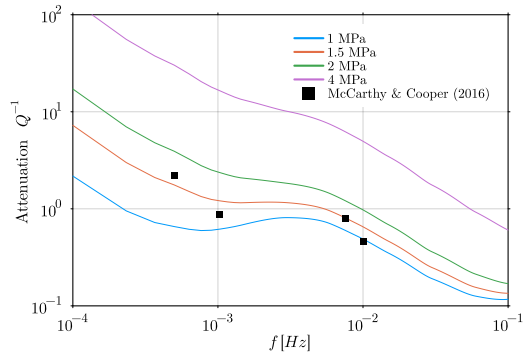

**Fig. 6.** Attenuation spectrum from the MCMC model for ice, for several median stress amplitudes on a log-log scale. Blue curve is for $\sigma_m = 1$ MPa and $\sigma_0 = 0.17$ MPa, corresponding to the experiments of McCarthy et al. (61). The red curve is $\sigma_m = 1.5$ MPa and $\sigma_0 = 0.5$ MPa; green curve is $\sigma_m = 2$ MPa and $\sigma_0 = 0.9$ MPa; and magenta is $\sigma_m = 4$ MPa and $\sigma_0 = 1.5$ MPa. The black squares are the coarse grain-size attenuation data from (61)

undergoes minimal variation, and the Debye peak becomes almost indiscernible. Moreover, the difference in attenuation observed between the 2 MPa and 4 MPa curves significantly exceeds the variation between 1 MPa and 2 MPa, underscoring the nonlinear nature of the phenomenon. It is also noteworthy that only the 1 MPa curve exhibits a negative slope, around the frequency of $10^{-3}$ Hz, whereas the curves corresponding to higher stresses exhibit positive monotonic behavior.

The results in Figure 6 show deviations from the attenuation values measured by McCarthy et al. (61). Several factors may account for these differences. First, the parameter values used in our numerical experiments were optimized based on dislocation creep data. However, the attenuation data from McCarthy et al. (61) pertains to steady-state creep within the GBS regime, occurring near the transition to dislocation creep. This discrepancy in experimental conditions could partially explain the observed disparities in absolute values. The stress exponent $n$ used in the numerical model is 4, whereas it might be lower, closer to 3, when GBS and dislocation creep contribute equally to the strain rate. Another contributing factor to the discrepancy is the difference in experimental environments. McCarthy et al. (61) conducted their experiments at ambient pressure with a maximum stress of 1 MPa, while the numerical model was optimized using data from experiments performed under confining pressure, often at much higher stress levels. The absence of confining pressure in the original experiments could lead to cracking in the samples, which dissipates energy and results in higher attenuation values. Additionally, low frequencies can induce significant strains, potentially leading to microstructural evolution in the samples; thus, the assumption of a constant microstructure might not be valid at frequencies below $10^{-3}$ Hz. Alignment of attenuation values could be achieved by optimizing the numerical model using experimental attenuation data. Employing the same Bayesian inference method could highlight differences in parameter values that the process converges upon, potentially allowing for a more accurate reproduction of the experimental attenuation spectrum. Despite these incogruencies introduced by the extrapolative approach, the resultant patterns and trends remain consistent with the experimental investigations. Figure 6 demonstrates a clear 'Debye peak' and a high-temperature background, expected from ice at these frequencies, and also illustrates a nonlinear viscoelastic property through amplitude dependence of the spectrum. These results suggest that the model effectively reproduces nonlinear viscoelastic properties of ice undergoing deformation via dislocation-based mechanisms in both time and frequency domains.

**Discussion & Conclusion.** Two main outcomes are presented in this study: first, a unified and versatile nonlinear viscoelastic model for geological materials, and second, the ability to use the results of MCMC simulations to predict other types of experiments. The second outcome is particularly important for both modeling and experimental design. Having a unique model, backed by observed microstructure in the lab, that may be used for large-scale models in geodynamics increases the confidence in large-scale predictions. From an experimental design viewpoint, this model can be used to test hypotheses before conducting experiments. Attenuation experiments in the dislocation creep regime are very difficult to perform due to limitations of existing apparatus. The model presented here was used to create an attenuation spectrum at elevated stress levels that can be verified by future experiments (Figure 6).

Moreover, the optimized model can be used for sensitivity analysis, exploring the effect of individual parameters on deformations or energy dissipation. Comparing the posterior convergences between the different parameters provides a physical understanding of the relative importance of each mechanism. For example, if the posteriors of the critical strains of steady-state grain size and CPO are different, i.e., one remains more uniform and one shows strong convergence into a normal distribution, it might indicate that one mechanism was more important in the deformation than the other, even though both were happening simultaneously.

While the data used in this study were from dislocation creep experiments, this type of model can work with different stress and grain size exponents to fit other types of data. It is important to test the model with other experiments to verify its applicability in different scenarios. The relatively large number of parameters in the model is beneficial but should also be considered with care. The benefit is enabling the usage of the model with various datasets, but reaching a good fit has a high probability if the parameter space of the priors is too large. Therefore, it is important to use the model with more datasets and provide the narrowest constraints possible for each parameter.

The formulation of our model includes a single Kelvin-Voigt element, corresponding to one anelastic relaxation time ($\tau_K = \eta_K / E_K$). Other viscoelastic models, such as the Andrade model, include an infinite series of dashpots and springs in the anelastic circuit to broaden the relaxation spectrum and better fit laboratory data (62–64). In our formulation, the nonlinear dashpot, along with the dynamic evolution of the microstructure, works to increase the relaxation times of the transient phase beyond the singular Kelvin-Voigt component. Conceptually, the nonlinear dependence of the dashpot on current stress levels and microstructure forms a relaxation time related to that specific state, continuously changing as the sample deforms.

Geodynamics data are limited and often rely on rare large-scale events such as earthquakes and volcanic eruptions.

Consequently, laboratory experiments remain crucial for advancing understanding in the field. Bridging the gap between the small scales of laboratory experiments and the large-scale events of geodynamic phenomena requires a specialized framework. The model proposed in this study, coupled with the MCMC approach, offers an effective framework to integrate large-scale observations with laboratory-generated data. The parameters in our model can be applied to both small and large-scale observations, enabling predictions on both scales. We hope this study will prove valuable for both experimentalists and geodynamics modelers in future research.

## Materials and Methods

**Bayesian Inference and MCMC.** The parameters of the nonlinear Burgers model were optimized using Bayesian inference (65). In this framework, we defined the state vector of our system as $\mathbf{x} = [\sigma, \varepsilon_K, \varepsilon_{M_e}, \varepsilon_{M_d}, d, F]^\top$, and the parameter space as $\theta = [E_M, E_K, \eta_K, \beta, Q_M, n, p, d_{ss}, F_{ss}, \varepsilon_{c1}, \varepsilon_{c2}]^\top$. Given a set of observations (data $\mathcal{D}$), the model is expected to represent it with an added error $\epsilon$: $\mathcal{D} = \mathcal{M}(\mathbf{x}, \theta) + \epsilon$. The posterior probability we estimated for each experiment is $P(\theta|\mathcal{D}, \mathcal{M})$, i.e., the probability of the parameter values conditioned on our experimental data and model. We inferred this posterior probability using Bayes' theorem and Hamiltonian Monte Carlo (HMC) sampling methods.

HMC is part of a large family of algorithms known as Markov Chain Monte Carlo (MCMC). In this stochastic modeling approach, the posterior distribution is the stationary distribution of the model. By using Markov chains to sample the parameter space, the resulting histograms of the sampling process represent the posteriors. Due to their Monte Carlo nature, many MCMC algorithms require a very large number of samples to build the posterior and usually have a 'burn-in' practice, where thousands of the initial iterations are discarded (e.g., Metropolis-Hastings and Gibbs sampling). HMC algorithms are designed to use gradient information on the target distribution to improve the rate of convergence, by using Hamiltonian dynamics (66, 67). However, HMC requires the tuning of two hyperparameters, which is usually challenging.

The sampling method we used is the No-U-Turn Sampler (NUTS), which is a variation of HMC that dynamically adapts the hyperparameters recursively (68). This sampling method enabled us to reach relatively fast convergence without a burn-in phase, and it is detailed in *SI Appendix*, MCMC Sampling Algorithm.

**Ice Data and Priors.** The entire dataset used for the ice experiments is from (58). For the Bayesian inference, we set up the prior assumptions for the parameters as detailed in Table 1. The model is highly sensitive to the values of the material parameter $\beta$ and the activation energy $Q_M$, both of which are poorly constrained for ice. Therefore, we decided to narrow the distribution of $\beta$ to a range of values in line with theory and previous observations while allowing the activation energy to have a wider range. Without constraining their values, the sampling algorithm can get 'stuck' in areas of the statistical manifold that are far from the observations, preventing convergence.

The parameters related to the Kelvin-Voigt circuit, $E_K$ and $\eta_K$ were given a wide uniform distribution. This is because these parameters are unique to the Burgers model and currently do not have any record of values for ice in dislocation creep. In addition, the steady state value of the CPO factor, $F_{ss}$, was given a wide range uniform distribution as a prior, this is because it is part of a parametrization with no previous values on record. The samples did go through dynamic recrystallization and the evolution of CPO was observed, so we expected $F_{ss}$ values to be less than 1 after the simulations.

**Olivine Data and Priors.** The viscoelastic properties of olivine have likely been studied more than those of any other geological material (e.g. 71–74). Therefore, most parameters of the model are well constrained for the olivine runs, and they are presented in Table 2.

**Table 1. Priors used for MCMC runs on data from constant strain rate experiments on ice**

| Parameter & Distribution | Details & Reference |
| --- | --- |
| $\eta_K \sim \mathcal{U}[10^9, 5 \times 10^{11}]$ [Pa · s] | Wide range |
| $E_K \sim \mathcal{U}[5 \times 10^8, 10^{10}]$ [Pa] | Wide range |
| $E_M \sim \mathcal{N}(\mu = 4, \sigma^2 = 2)$ [GPa] | (69) |
| $d_{ss} \sim \mathcal{U}[10, 200]$ [µm] | (58) |
| $F_{ss} \sim \mathcal{U}[0.1, 0.9]$ | Wide range |
| $\varepsilon_{c1} \sim \mathcal{U}[0.01, 0.1]$ | (58) |
| $\varepsilon_{c2} \sim \mathcal{U}[0.01, 0.5]$ | (58) |
| $p \sim \mathcal{U}[0.1, 0.7]$ | (58) |
| $n \sim \mathcal{U}[2.7, 4.5]$ | (33) |
| $Q_M \sim \mathcal{N}(\mu = 65, \sigma^2 = 2)$ [kJ/mol] | (70) |
| $\beta \sim \mathcal{N}(\mu = 7.29 \times 10^{-4}, \sigma^2 = 10^{-6})$ [J$^{-1}$] | (33) |

**Table 2. Priors used for MCMC runs on data from the attenuation experiments on olivine**

| Parameter & Distribution | Details & Reference |
| --- | --- |
| $\eta_K \sim \mathcal{U}[5 \times 10^8, 10^{12}]$ [Pa · s] | Wide range |
| $E_K \sim \mathcal{U}[7 \times 10^8, 9 \times 10^{10}]$ [Pa] | Wide range |
| $E_M \sim \mathcal{N}(\mu = 200, \sigma^2 = 10)$ [GPa] | (75) |
| $\varepsilon_{c1} \sim \mathcal{U}[0, 1]$ | (19) |
| $p \sim \mathcal{U}[0.2, 1.5]$ | (54) |
| $n \sim \mathcal{U}[2.5, 3.9]$ | (54) |
| $Q_M \sim \mathcal{N}(\mu = 445, \sigma^2 = 20)$ [kJ/mol] | (54) |
| $\beta \sim \mathcal{N}(\mu = -8.6 \times 10^{-5}, \sigma^2 = 10^{-5})$ [J$^{-1}$] | (54) |

We used data from a forced oscillation experiment performed in the Deformation-DIA apparatus, in which the differential stress and strain were measured in-situ using synchrotron energy-dispersive X-ray diffraction and radiography, respectively. The experiment featured a bias stress resulting in a temporally variable background strain rate. We corrected the strain data for this background strain rate by fitting a sinusoid combined with a polynomial trend to the strain data and subtracting the polynomial trend from the fit. Then, we determined the strain rate associated with the forced oscillations by differentiation of the sinusoidal fit. This strain rate was used by the model to calculate the stress. The details of this process, along with the original data are presented in *SI Appendix*, Olivine.

Initial analysis of the microstructures before and after forced oscillations indicated negligible CPO evolution. The grain-size evolution was parametrized using the recrystallized grain size piezomoeter proposed by (76):

$$\dot{d} = \frac{\dot{\varepsilon}\left(0.015 \times \sigma^{-1.33} - d\right)}{\varepsilon_{c1}} \quad [5]$$

Additional microstructural analysis of the sample performed after the MCMC runs indicated an almost negligible grain-size evolution as well. This conclusion is further supported by the non-convergence of the grain-size evolution parameter (Fig. S5B).

Both the olivine and ice runs consisted of 5 MCMC chains, each generating 3,000 accepted samples to assess the posterior.

**ACKNOWLEDGMENTS.** We wish to thank Diede Hein for generously sharing the data from the attenuation experiments on olivine and for his assistance in constraining the priors and formulating the olivine model. We also extend our gratitude to Travis Hager for his comments and suggestions regarding the ice model and data.